# Hot carrier solar cells by adiabatic cooling

Tom Markvart

Engineering Sciences, University of Southampton, Southampton SO17 1BJ, UK, and Centre for Advanced Photovoltaics, Czech Technical University, 166 36 Prague 6, Czech Republic

**Abstract**

Hot carrier solar cell is proposed where charge carriers are cooled adiabatically in the charge transport layers adjoining the absorber. The device resembles an ideal thermoelectric converter where thermopower, and therefore also carrier entropy, are maintained constant during cooling from the temperature attained in the absorber to the temperature at contacts.

Hot carrier solar cells represent an attractive opportunity for increasing the cell efficiency above the Shockley-Queisser limit.[1] Proposed originally by Ross and Nozik[2] in 1982, the area has become an active area of research to the present day. [3,4,5] Most progress has been made trying to maintain the injected carriers at an elevated temperature for extended periods of time. [6,7,8] Less successful have been attempts to implement these materials into a device. Indeed, Ross and Nozik's idea to cool carriers in energy selective contacts has not been subjected to a serious theoretical examination (see, however, Ref. 9), and few experimental studies confirm efficacity of this approach. In this paper, an alternative method of adiabatic carrier cooling is suggested, based on thermodynamics.

The proposed hot carrier solar cell is shown schematically in Fig. 1a. The device bears a close resemblance to a standard thin-film or perovskite solar cell (see, for example, refs. 10,11) with the additional feature of a controlled doping profile in the charge transport layers, to be discussed below. Photons are absorbed in the absorber, sandwiched between n and p charge transport layers, with minimum absorption in the wide-bandgap front charge transport layer. By virtue of different work function of the two charge transport layers, the absorber contains a strong electric field which separates the photogenerated electrons and holes (Fig. 1b). The carriers are separated in time $\tau_{sep} = d/v$, where $d$ is the absorber thickness, $v = \mu \mathscr{E}$ is the carrier velocity, $\mu$ is the mobility and $\mathscr{E}$ is the electric field. If the absorber is made from material that maintains carriers hot over time $t > \tau_{sep}$, carriers separated to the edges of the absorber will retain an elevated temperature $T_h$.

A quick estimate is useful to illustrate the feasibility of this process. A typical absorber thickness is of the order of micrometers. Taking a typical carrier velocity $v \sim 10^7$ cm/sec,[12] we arrive at a time $\tau_{sep}$ of the order of 10 psec, already achieved in some research materials,[13] putting separation of hot carriers in a junction/absorber within reach.

The challenge is to cool down these carriers in the charge transport layers to ambient temperature $T_o$ at contacts, without losing energy by thermalization. In thermodynamic terms this means that the cooling has to take place adiabatically, by keeping the carrier entropy $s$ constant. A standard result, using the Sackur-Tetrode equation,[14] gives

$$s = k_B \ln\left(\frac{T^{3/2}}{n}\right) + \text{constant} \qquad (1)$$

where $n$ is the carrier density. An alternative argument based on carrier thermopower (Seebeck coefficient) $S = \pm s/q$, where $q$ is the elementary charge, leads to the same result. With temperature profile such as in Fig. 1b, a doping profile

$$n = \text{constant} \times T^{3/2} \qquad (2)$$

in the charge transport layer (Figs 1c) will keep the carrier entropy constant, achieving adiabatic cooling. In particular, the required doping at the absorber and contact edges of the cooling / charge transport layers satisfies

$$\frac{n(absorber)}{n(contact)} = \left(\frac{T_h}{T_o}\right)^{3/2} \qquad (3)$$

The corresponding band diagram of such a hot carrier solar cell at equilibrium is shown in Fig. 1d.

The maximum achievable open-circuit voltage can estimated by using the relationship between thermopower and the electrochemical potential $\mu_{ec}$,[15]

$$\nabla \mu_{ec} = -qS\nabla T = \pm s \nabla T \qquad (4)$$

with the – sign for holes and + sign for electrons. Integrating (4) over the cell from the n to the p contact we obtain

$$qV_{oc} = -\int_{n\,contact}^{p\,contact} s\,\nabla T = (s_n + s_p)(T_h - T_o) \qquad (5)$$

bearing in mind that the electron ($s_n$) and hole ($s_p$) entropies are constant. Now, $(s_n + s_p)T_h$ is the combined heat $Q$ per carrier that flows from the absorber into the charge transport layers which, if the absorber is made of a perfect hot-carrier material, is equal to the energy absorbed. Therefore, for an ideal hot carrier cell,

$$qV_{oc} = (s_n + s_p)T_h\left(1 - \frac{T_o}{T_h}\right) = Q\left(1 - \frac{T_o}{T_h}\right) \qquad (6)$$

Hence, the voltage efficiency is equal to

$$\eta_V = \frac{qV_{oc}}{Q} = \left(1 - \frac{T_o}{T_h}\right) \qquad (7)$$

- in other words, the hot carrier cell produces voltage with Carnot efficiency, avoiding thermalization losses. A similar idea to improve generated voltage could be applied to thermoelectric converters.

The power conversion efficiency which follows from (7) is shown in Fig. 2 as a function of the semiconductor bandgap, compared with the maximum efficiency attainable by a single-junction cell, as given by the Shockley-Queisser detailed balance.[1] In the absence of thermalization, the efficiency of hot-carrier cells reaches a maximum 86.4% at zero bandgap, limited only by the reversible loss (6) and losses implied by finite-time thermodynamics to produce power rather than energy, and expressed by the fill factor.[16] Even for standard photovoltaic materials with bandgaps 1– 1.8 eV, the hot carrier cell efficiency exceeds the Shockley-Queisser limit by a factor of 1.4 – 1.9 under concentrated sunlight, and by a factor of 1.5 – 2.2 under one sun illumination – a worthwhile goal to aim for.

This theoretical gain is mitigated by efficiency losses encountered in practice. In addition to challenges represented by the search for suitable hot carrier materials, there are standard optical and resistive losses, as well as losses by nonradiative recombination which are well researched in conventional solar cells. In addition, another loss should be taken into account: losses implied by photons transported to the solar cell by radiative transfer rather than heat transfer. As a result, near-equilibrium solar photons at temperature $T_S$ ~ 6000 K generate carriers with the same energy but away from equilibrium. An equilibrium at a different temperature $T_h$ is reached via carrier-carrier scattering within less that some 100 fsec, faster than scattering with phonons.[13] This process, however, is irreversible and increases entropy, resulting in energy (and hence voltage) loss, although there is potential for reducing this loss by distributing excited carriers between several bands.[17]

**Acknowledgement**
This work was supported by project "Energy Conversion and Storage", grant no. CZ.02.01.01/00/22_008/0004617, programme Johannes Amos Commenius, Excellent Research.

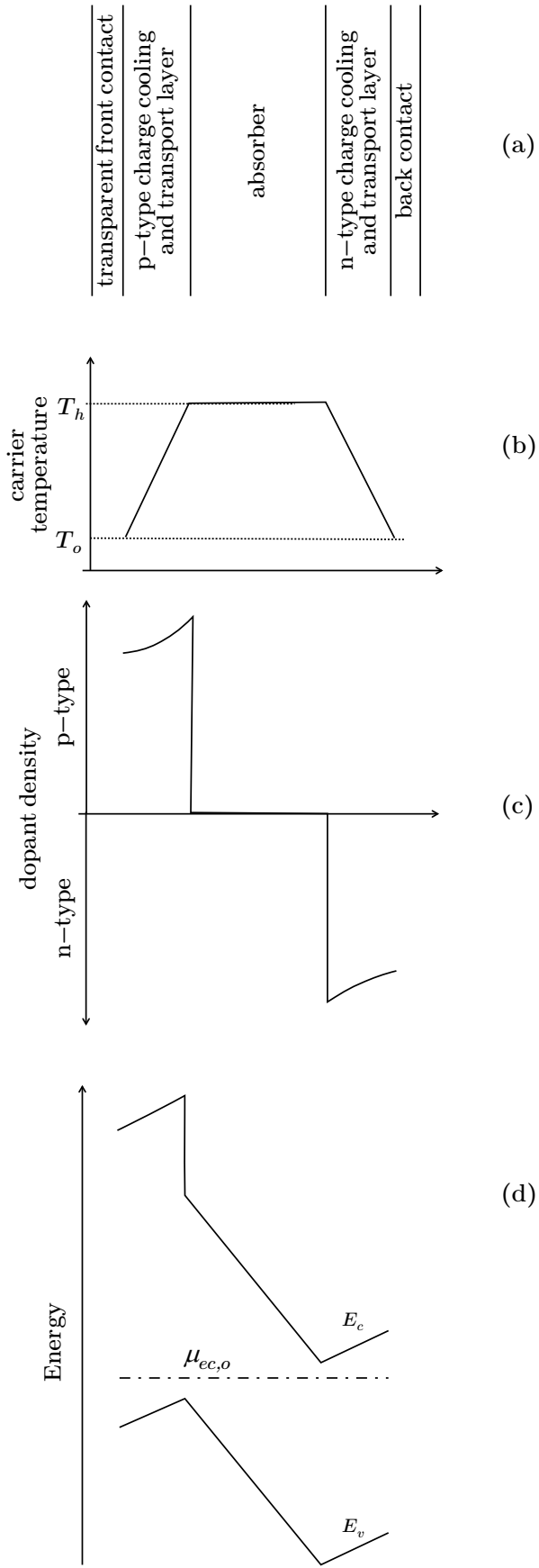

Fig. 1 (a) A schematic diagram of the hot carrier solar cell with adiabatic cooling in the charge transport layers. (b) The carrier temperature profile in operation. (c) The doping profile. (d) The band diagram at equilibrium.

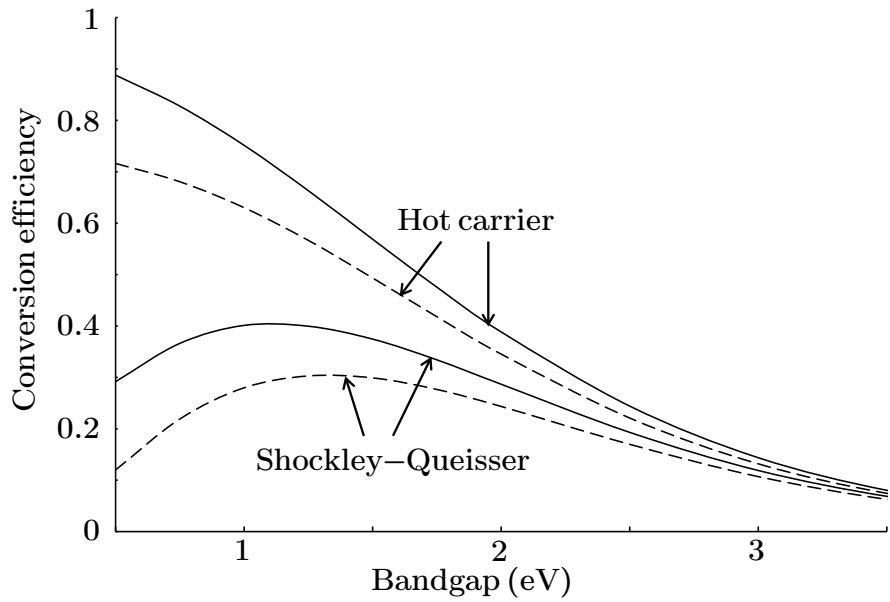

Fig. 2. The efficiency of hot carrier solar cell as a function of the bandgap, at maximum concentration of sunlight (full lines) and at one-sun illumination (dashed lines).